\documentstyle[aps,prbbib,twocolumn,epsf]{revtex}

\begin{document}
\draft
\title{Non-adiabatic charge pump: an exact solution}

\author{Baigeng Wang$^1$, Jian Wang$^{1,2,a}$, and Hong Guo$^3$}
\address{1. Department of Physics, The University of Hong Kong, 
Pokfulam Road, Hong Kong, China\\
2. Institute of Solid State Physics, Chinese Academy of Sciences,
Hefei, Anhui, China\\
3. Department of Physics, McGill University, Montreal, Quebec, 
Canada H3A 2T8}
\maketitle

\begin{abstract}
We derived a general and exact expression of current for quantum parametric 
charge pumps in the non-adiabatic regime at finite pumping frequency and finite 
driving amplitude. The non-perturbative theory predicts a remarkable plateau
structure in the pumped current due to multi-photon assisted processes in
a double-barrier quantum well pump involving only a {\it single} pumping
potential.  It also predicts a current reversal as the resonant level of the
pump crosses the Fermi energy of the leads.  
\end{abstract}


Parametric charge pump is a device that drives the flow of a DC electric
current by a time dependent variation of {\it two} or more device parameters. 
The quantum parametric pump has received considerable attention both 
experimentally and theoretically\cite{brouwer,switkes,zhou,shutenko,wei1,levinson,avron,buttiker1,vavilov1,wang1,wbg1,buttiker2}.
A very useful and clear transport theory on quantum parametric pump was 
provided by Brouwer\cite{brouwer} for the adiabatic regime, {\it i.e.} in the
$\omega\rightarrow 0$ limit, where $\omega$ is the frequency of driving force 
of the pump. Since then there have been several attempts devoted towards the 
understanding of parametric pumping at finite frequency, {\it i.e.} for the
more general non-adiabatic regime\cite{buttiker1,vavilov1,wang1}.
These theories\cite{buttiker1,vavilov1,wang1} were all based perturbative 
schemes either in terms of pumping frequency or in pumping amplitude. Recently, 
a Floquet scattering matrix theory was developed\cite{buttiker2} and an exact 
solution obtained for an oscillating double-barrier pump structure. Despite 
the progress, so far a general and non-perturbative theory for quantum 
parametric pump has not been reported. Such a theory will be very important
to establish a general and unambiguous physical picture of quantum charge
pumps for the non-adiabatic regime, and provide new features not accessible 
by perturbative theories. It is the purpose of this Letter to present the 
exact solution for the current delivered by a quantum parametric pump at 
{\it both} finite driving frequency and amplitude. 

Our theory is based on Keldysh non-equilibrium Green's functions
(NEGF)\cite{wbg1} and is non-perturbative. It reproduces not only all 
previous 
theories in their perspective limits, but also reveals qualitative new and 
interesting transport properties of quantum parametric pump. Our theory is 
also applicable to the case where the driving potential is in the interior 
of the pump. Because of the generality of our theory, it is now possible 
to investigate situations beyond the established physics of adiabatic 
pump. As an example, we investigate a peculiar case of a quantum-well charge 
pump driven by only a {\it single} parameter: such a pump is only possible 
in the non-adiabatic regime. In particular, if the pumping potential is 
located at the barriers of the quantum well, the pumped current exhibits 
remarkable plateaus due to a multi-photon assisted pumping process. The width 
of the plateaus is determined by $\hbar \omega$ while their height vary 
depending on the pumping amplitude and the coupling strength between the 
pump and the leads. By tuning a gate voltage, the resonant level of the 
quantum well crosses Fermi energy of the leads and a pumped current reversal 
is observed. On the other hand, if the pumping potential is at the center 
point of the quantum well, due to the spatial symmetry the pumped current 
vanishes exactly. Moving away from this symmetric point, the pumped current 
increases rapidly by several orders of magnitude and similar plateaus 
in pumped current are observed.

Now we derive the pumped current in the non-adiabatic regime. Assuming that 
the time-dependent driving potential has the following form 
$H'(t_1)=(V/2)\exp(i\omega t_1)+(V^*/2)\exp(-i\omega t_1)$ where $V$ is the 
effective pumping potential profile. Neglecting interaction between electrons 
in the ideal leads labeled by $L$ and $R$, we start from the expression of
time averaged current\cite{jauho} ($\hbar=1$) in standard NEGF theory,
\begin{eqnarray}
I_{\alpha}&=&-\frac{q}{\tau}\int_0^\tau dt \int dt_{1}
Tr[{\bf G}^{r}(t,t_{1}){\bf \Sigma}_{\alpha}^{<}(t_{1},t) \nonumber \\
&+&{\bf G}^{<}(t,t_{1}){\bf \Sigma}
_{\alpha}^{a}(t_{1},t)+c.c.]
\label{X4average}
\end{eqnarray}
where $\tau=2\pi/\omega$ is the period of pumping cycle. Using the following 
double time Fourier transform
\begin{equation}
G^{\gamma}(t_1, t_2) = \int \frac{dE_1}{2\pi} \frac{dE_2}{2\pi} 
e^{-i E_1  t_1 + i E_2 t_2} G^{\gamma}(E_1, E_2)
\label{X2fourier2}
\end{equation}
with $\gamma=r,a,<$, Eq.(\ref{X4average}) becomes,
\begin{eqnarray}
&&I_\alpha = -\frac{q}{2N\tau} \int \frac{dE_1}{2\pi} \frac{dE_2}{2\pi} 
{\rm Tr} \{ [G^r (E_1,E_2) -G^a(E_1,E_2)] 
\nonumber \\
&&\Sigma^<_\alpha(E_2,E_1) + G^<(E_1,E_2) [\Sigma^a_\alpha(E_2,E_1)
-\Sigma^r_\alpha(E_2,E_1)]\} \nonumber \\
&&\equiv -\frac{q}{8N\tau \pi^2} {\rm Tr} [({\bf G}^r-{\bf G}^a) 
{\bf \Sigma}^<_L + {\bf G}^< ({\bf \Sigma}^a_L - {\bf \Sigma}^r_L)]
\label{new}
\end{eqnarray}
where we have extended the integration range for $dt$ in Eq.(\ref{X4average})
to $[-N\tau,N\tau]$ with $N->\infty$. Note that in Eq.(\ref{new}) the matrix 
${\bf G}^\gamma$ has element $G^\gamma(E_1,E_2)$. It is also easy to show that 
${\bf G}^r - {\bf G}^a = -i {\bf G}^r {\bf \Gamma} {\bf G}^a$.
Using the fact that ${\bf G}^< = {\bf G}^r {\bf \Sigma}^< {\bf G}^a$
and $\Sigma^\gamma(E_2,E_1) = 2\pi \delta(E_1-E_2) 
\Sigma^\gamma(E_1)$ (no external bias), we obtain 
\begin{eqnarray}
I_\alpha &=& \frac{q}{2N\tau}\int \frac{dE_1}{2\pi} \frac{dE_2}{2\pi} 
{\rm Tr} \left[ \Gamma_\alpha(E_1) G^r(E_1,E_2) \right. \nonumber \\
&& \left. \Gamma(E_2) G^a(E_2,E_1) \right] (f(E_2)-f(E_1))
\label{X4general}
\end{eqnarray}
Using the identity 
\begin{equation}
\sum_E = 2N\tau \int dE/(2\pi)
\label{iden}
\end{equation}
we finally arrive at 
\begin{eqnarray}
I_\alpha &=& \frac{q}{(2N\tau)^2}\int \frac{dE}{2\pi} 
\sum_{n=-\infty}^{+\infty} 
{\rm Tr} \left[ \Gamma_\alpha(E) G^r(E,E+n\omega) \right. \nonumber \\
&& \left. \Gamma_n(E) G^a(E+n\omega,E) \right] (f_n(E)-f(E))
\label{final}
\end{eqnarray}
where $\Gamma_n(E)=\Gamma(E+n\omega)$ and $f_n(E)=f(E+n\omega)$. Now we 
proceed to calculate $G^r(E,E+n\omega)$. From the Dyson equation, we have 
\begin{equation}
G^r(t,t')=G^{0r}(t,t')+\int G^r(t,t_1) H'(t_1) G^{0r}(t_1,t') dt_1
\label{eq2}
\end{equation}
where $G^{0r}$ is the equilibrium Green's function. Taking a double-time 
Fourier transform of Eq.(\ref{eq2}), we find 
\begin{eqnarray}
G^r(E_1,E_2)&=&2\pi G^{0r}(E_1) \delta(E_1-E_2)\nonumber \\
&+&\int \frac{dE}{2\pi} G^r(E_1,E_2+E) H'(E) G^{0r}(E_2) \ .
\end{eqnarray}
Since $H'(E)= \pi [V \delta(E+\omega)+ V^*\delta(E-\omega)]$, we obtain
\begin{eqnarray}
&&G^r(E_1,E_2)=2\pi G^{0r}(E_1) \delta(E_1-E_2) \nonumber \\
&+&[G^r(E_1,E_2-\omega) V+ G^r(E_1,E_2+\omega) V^*] G^{0r}(E_2)/2\ .
\label{X4gr1}
\end{eqnarray}
Using this equation, the general expression of $G^r(E,E+n\omega)$ can
be obtained. To simplify notation, we use the following abbreviations 
$G^r_n(E) \equiv G^r(E,E+n\omega)$ and $G^{0r}_n(E) \equiv G^{0r}(E+n\omega)$.
We then have
\begin{eqnarray}
&& G^r_0 = 2\pi G^{0r}_0 \delta(0) + (G^r_1 V^* + G^r_{-1} V) G^{0r}_0/2
\nonumber \\
&& G^r_1 = (G^r_2 V^* + G^r_0 V) G^{0r}_1/2
\nonumber \\
&& G^r_{-1} = (G^r_0 V^* + G^r_{-2} V) G^{0r}_{-1}/2
\label{eq1}
\end{eqnarray}
If we restrict ourselves for single photon process only, we can neglect 
$G^r_2$ in Eq.(\ref{eq1}) and obtain 
\begin{equation}
G^r_0(E) = \frac{2\pi \delta(0)}{[G^{0r}_0(E)]^{-1}-\Sigma^r_1(E)}
\label{X4gr1a}
\end{equation}
where $2\pi\delta(0)=\int dE=2N\tau$. Here the self-energy 
$\Sigma^r_1(E) \equiv (V G^{0r}_1 V^*+ V^* G^{0r}_{-1} V)/4$.

The exact solution of $G^r_n$ by including all photon processes can be 
obtained by iterating the following equation obtained from Eq.(\ref{X4gr1}),
\begin{equation}
G^r_n = (G^r_{n+1} V^* + G^r_{n-1} V) G^{0r}_n/2
\end{equation}
we obtain
\begin{equation}
G^r_n = G^r_{n-1} V G^{0r}_n \frac{1}{2\alpha^r_n}
\label{grn}
\end{equation}
where $\alpha^r_n=[(a_n,\bar{a}_n),(a_{n+1},\bar{a}_{n+1}),
...]$ with $a_n=iV G^{0r}_{n+1}/2$ and 
$\bar{a}_n=iV^* G^{0r}_n/2$. The continued fraction 
$[(a_1,\bar{a}_1),(a_{2},\bar{a}_2),...]$ is defined as
\begin{equation}
[(a_1,\bar{a}_1),(a_{2},\bar{a}_2),...]
=1+a_1\frac{1}
{\displaystyle 1+a_2\frac{1}{\displaystyle
1+... }\bar{a}_2}\bar{a}_1
\end{equation}
From Eq.(\ref{grn}), we obtain $G^r_1=G^r_0 V G^{0r}_1/(2\alpha^r_1)$
and $G^r_{-1}=G^r_0 V^* G^{0r}_{-1}/(2\beta^r_{-1})$. Here 
$\beta^r_{-n}=[(b_{-n},\bar{b}_{-n}),(b_{-n-1},\bar{b}_{-n-1}),...]$ 
with $b_{-n}=i V^* G^{0r}_{-n-1}/2$ and $\bar{b}_{-n}=i V G^{0r}_{-n}/2$. 
Substituting the expressions of $G^r_{\pm 1}$ into Eq.(\ref{eq1}), we have 
\begin{equation}
G^r_0(E) = \frac{2\pi \delta(0)}{[G^{0r}_0(E)]^{-1}-\Sigma^r(E)}
\label{X4gr1b}
\end{equation}
with 
\begin{equation}
\Sigma^r= V G^{0r}_1 \frac{1}{4\alpha^r_1} V^*
+V^* G^{0r}_{-1} \frac{1}{4\beta^r_{-1}} V
\label{self}
\end{equation}
Once $G^r_0$ is obtained, $G^r_n$ can be calculated recursively from 
Eq.(\ref{grn}). Eqs.(\ref{final}), (\ref{grn}), (\ref{X4gr1b}) and 
(\ref{self}) form the central result of this paper\cite{foot2}.  
Note that $G^{0r}_n$ is the Green's function for the process of absorbing 
(emitting) n-photons whereas $G^r_n$ is the renormalized n-photon Green's 
function containing the contribution of multi-photon process. From 
Eq.(\ref{grn}), we see that $G^r_n$ is at least of the order of $V^n$. 
For this reason, as we will see below that in the resonant tunneling regime, 
the summation in Eq.(\ref{final}) converges rapidly even in the strong pumping
regime. Using the fact that ${\bf G}^r {\bf \Gamma} {\bf G}^a
={\bf G}^a {\bf \Gamma} {\bf G}^r$, it is easy to show $\sum_\alpha
I_\alpha=0$, {\it i.e.}, current conservation for parametric pumping.

Eq.(\ref{final}) reproduces all previously known results of parametric
pumping. We give two examples here. First, in Ref.\onlinecite{wbg1}, the 
pumped current is obtained using perturbation theory for finite pumping 
frequency up to the second order in pumping amplitude\cite{foot1}. This 
result is recovered by only keeping the terms $n=\pm 1$ in Eq.(\ref{final}) 
to obtain $G^r_1 = G^{0r}_0 V G^{0r}_1/2$. Hence Eq.(\ref{final}) becomes,
\begin{equation}
I_\alpha = q\int \frac{dE}{8\pi} \sum_{n=\pm 1} {\rm Tr} \left[
\Gamma_\alpha G^{0r}_0 V G^{0r}_n \Gamma_n G^{0a}_n V^* G^{0a}_0 \right] 
(f_n-f)
\label{wbg1}
\end{equation}
which is exactly the same as that of Ref.\onlinecite{wbg1,foot1}. Second, we
consider the well-studied adiabatic regime $\omega\rightarrow 0$ where
the instantaneous approximation for the Green's function is appropriate. 
We transform Eq.(\ref{final}) to the Wigner representation using 
\begin{equation}
G^{\gamma}(t_1, t_2) = \int \frac{dE}{2\pi} 
e^{-i E ( t_1 - t_2)} {\cal G}^{\gamma}(E, T)
\end{equation}
where $T=(t_1+t_2)/2$ and ${\cal G}^\gamma$ is the NEGF in the Wigner 
representation. We then have\cite{foot3}
\begin{equation}
G^r_n=\int dT {\cal G}^r(E,T) e^{-i n \omega T}
\label{wigner}
\end{equation}
Substituting Eq.(\ref{wigner}) into Eq.(\ref{final}) and keep only the
$O(\omega)$ term, we obtain
\begin{eqnarray}
&&I_\alpha = \frac{q}{(2N\tau)^2}\int \frac{dE}{2\pi} 
\int_{-\infty}^\infty dT dT' \sum_{n=-\infty}^{+\infty} n\omega ~ 
{\rm Tr} \left[ \Gamma_\alpha(E) \right. \nonumber \\
&& \left. {\cal G}^r(E,T) \Gamma(E) {\cal G}^a(E,T') \right] 
\exp(-i n \omega (T-T')) \partial_E f(E) 
\label{temp1}
\end{eqnarray}
Note that
\begin{eqnarray}
&&\int dT \sum_{n=-\infty}^{+\infty} n\omega ~ {\cal G}^r(E,T) 
\exp(-i n \omega (T-T')) \nonumber \\
&&= -2i N\tau ~ \partial_{T'} {\cal G}^r(E,T') 
\end{eqnarray}
where we have used Eq.(\ref{iden}).  Eq.(\ref{temp1}) now becomes,
\begin{eqnarray}
I_\alpha = \frac{-iq}{\tau} \int_0^\tau dT \int \frac{dE}{2\pi} 
{\rm Tr} \left[ \Gamma_\alpha \partial_T {\cal G}^r(T)
\Gamma {\cal G}^a(T) \right] \partial_E f\ , 
\label{temp2}
\end{eqnarray}
this is exactly the Brouwer's formula\cite{brouwer,buttiker2} for adiabatic
pumping.

In the following we calculate pumped current using Eq.(\ref{final}) for 
a symmetric double $\delta$-barrier structure given by
$U(x)=V_0 \delta (x+a)+V_0 \delta (x-a)$. For this system the Green's function 
$G(x,x')$ can be calculated exactly\cite{yip}. In the adiabatic theory, two
pumping potentials are needed in order to give a nonzero pumping current. 
At finite frequency, a {\it single} pumping potential was reported to be 
enough to pump a current\cite{wbg1} which is peculiar, and our exact 
non-perturbative theory derived above allows us to investigate this
situation clearly and unambiguously. We chose the single pumping potential 
to be sinusoidal $V(x,t)=V_p \cos(\omega t) \delta(x-x_0)$, and we
calculate pumped current from the left lead at zero temperature and set 
$V_0=200$ unless specified otherwise. A gate voltage $v_g$ is applied to the 
double barrier structure so that the resonant single particle energy level
is controlled by it. We fix the Fermi level of leads in line with the resonant 
level at $v_g=0$. Finally the unit is set by $\hbar=2m=q=2a=1$. For GaAs
material with well width $a=1000\AA$, the energy unit is $E=0.056meV$
which corresponds to $\omega=13.2$ GHz. The unit for pumped current is
$5\times 10^{-10}$ A.

Fig.1 shows the pumped current versus the gate voltage in the strong
pumping regime for $x_0=-a$, $V_p=160$, and $\omega=1$. Following observations 
are in order. (1) The pumped current displays a series of remarkable 
plateaus. The width of a plateau is equal to $\hbar\omega$ whereas the height
of the plateau depends on $V_p$ and $V_0$ in a nonlinear fashion. 
(2) The pumped current reverses its direction when the resonant level across
the Fermi level of the lead, {\it i.e.}, when $v_g$ changes sign. As a result, 
we see from Fig.1 that the pumped current is quenched near $v_g=0$. (3) The
height of the current plateaus for negative $v_g$ is considerablely
larger than those of positive $v_g$. To understand these results, we rewrite 
Eq.(\ref{final}) as
\begin{equation}
I_L = q \int \frac{dE}{2\pi} \sum_{n=1}^{+\infty} 
F_n(E) (f_n(E)-f(E))
\label{final1}
\end{equation}
with 
\begin{eqnarray}
F_n(E)&=& \frac{q}{(2N\tau)^2} {\rm Tr} \left[ \Gamma_L(E) 
G^r(E,E+n\omega) \Gamma_n(E) \right. \nonumber \\
&& \left. G^a(E+n\omega,E) - \Gamma_{L n}(E) G^r(E+n\omega,E) 
\right. \nonumber \\
&& \left. \Gamma(E) G^a(E,E+n\omega) \right] 
\end{eqnarray}
We notice that kernel $F_n$ consists of two terms: one due to photon
absorption process\cite{foot6} and the other to photon emission process. 
These two contributions differ by a sign indicating the competition 
between the photon absorption and emission processes similar to that
discussed in Ref.\onlinecite{wbg1}. In Fig.2 we plot the integrand of 
Eq.(\ref{final1}), 
$F_n(E)$, for $n=1,2,3$ (solid line, dotted line, dashed line, 
respectively). The sidebands due to photon absorption (when $v_g<0$) and 
photon emission are clearly seen (for an n-photon emission process, the 
energy has been shifted by $n\omega$ due to the transformation 
from Eq.(\ref{final}) to Eq.(\ref{final1})). To obtain pumped current,
one integrates these sidebands over energy. For the n-th sideband ($n>0$), 
the integral range is $[v_g-n\omega, ~v_g]$: it is therefore clear that the 
plateau structure is a direct result of multi-photon assisted processes.
The positive current is due to photon absorption processes and the
current reverses its sign if photon emission process dominates. The behavior 
of current reversal has been observed before for pumping in Carbon 
nanotubes\cite{wei2}, charge quantization\cite{aharony}, and heat current 
generated during the pumping\cite{buttiker2}, although from completely 
different origins. Due to the energy dependence of coupling between the 
pump and the leads, the sideband is asymmetric with a larger peak for 
photon absorption process. As a result, the height of the current plateau 
is larger for photon absorption process. Our numerical plots show that this 
plateau structure persists for different frequencies and pumping amplitudes. 
Our analysis show that there are two effects which affect the current 
plateau structure. The first is barrier heights of the quantum well:
current plateaus can only be observed in the strong tunneling regime 
and they disappear for low barrier height. The second is temperature:
the plateau structure is rounded off at finite temperature and destroyed 
when temperature reaches $\sim 200 mK$ for $\omega=10$GHz. 

When the pumping potential is inside of the double-barrier structure, we 
found a similar plateau structure due to the same photon assisted processes. 
In Fig.3, the pumped current as a function of pumping position $x_0$ is
plotted at a fixed $v_g=-0.67$ for a much smaller pumping amplitude $v_p=6$.
We observe that the pumped current $I_p$ is {\it antisymmetric} about the 
center of the pump. As expected for a symmetric structure, $I_p$
vanishes identically at $x_0=0$. At $x_0=-0.5$, we found $I_p=2\times 10^{-6}$. 
Fig.3 shows that $I_p$ can increase by three orders of magnitude by varying 
the position of the pumping potential away from the barriers.

In summary, we have derived a non-perturbative and exact current expression 
for parametric pumping at finite frequency and finite pumping amplitude. 
This theory can also account for the case of pumping potential located in 
the {\it interior} of the scattering region. Our theory reproduces all
previously known results in both adiabatic regime and in perturbative theory.
For a double-barrier quantum well pump, we predicted current plateaus to
appear due to multi-photon assisted processes with the plateau width 
given by the pumping frequency. As the gate voltage is varied and the
resonance level sweeps through the Fermi energy of leads, the pumped current 
is found to reverse its direction. 

\section*{Acknowledgments}
We gratefully acknowledge support by a RGC grant from the SAR Government of 
Hong Kong under grant number HKU 7113/02P and from NSERC of Canada and FCAR 
of Quebec (H.G).

\bigskip

\noindent{$^{a)}$ Electronic mail: jianwang@hkusub.hku.hk}

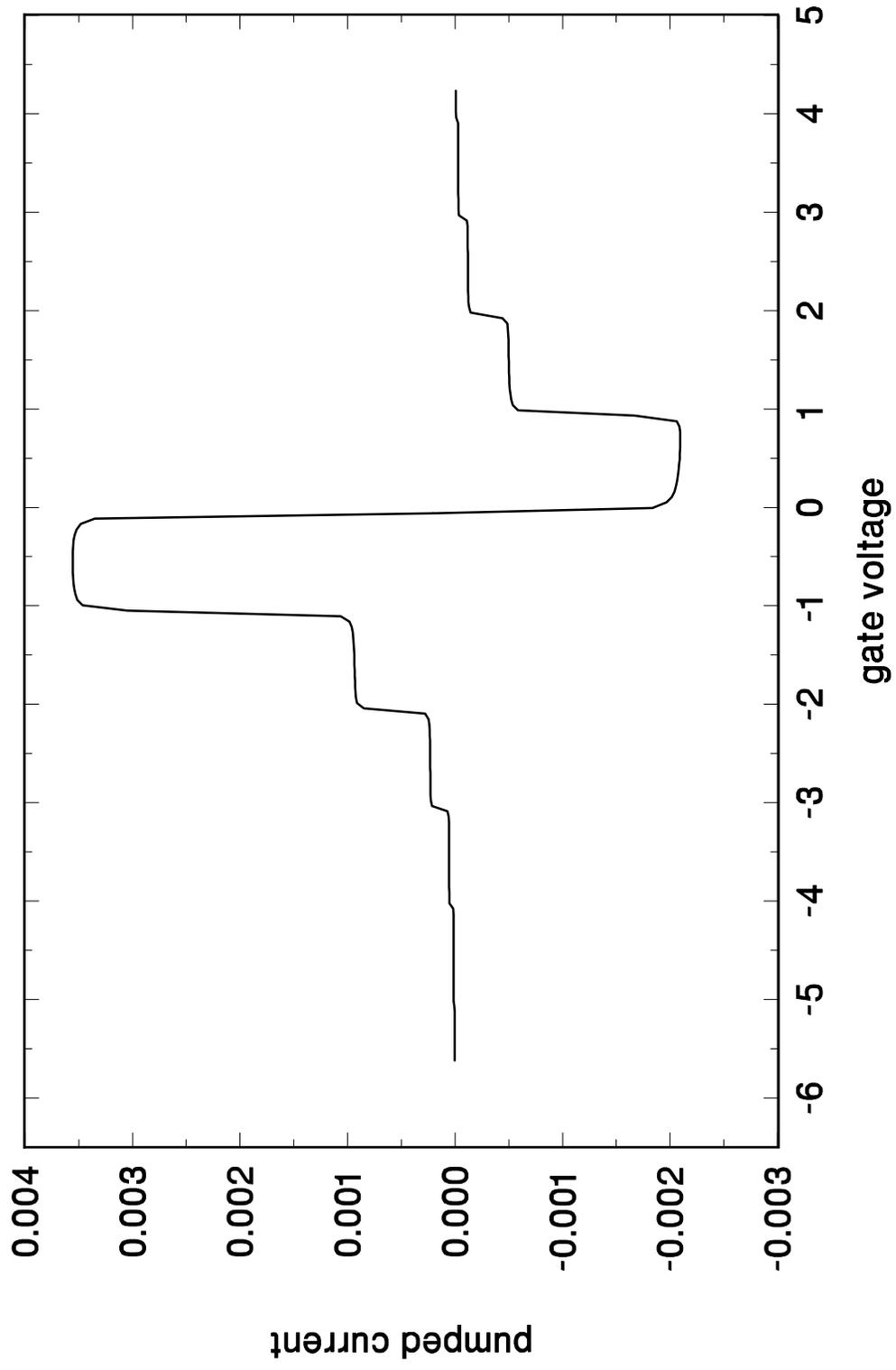
\begin{figure}
\caption{
The pumped current versus the gate voltage for $x_0=-a$, $\omega=1$, and
$v_p=160$.
}
\end{figure}

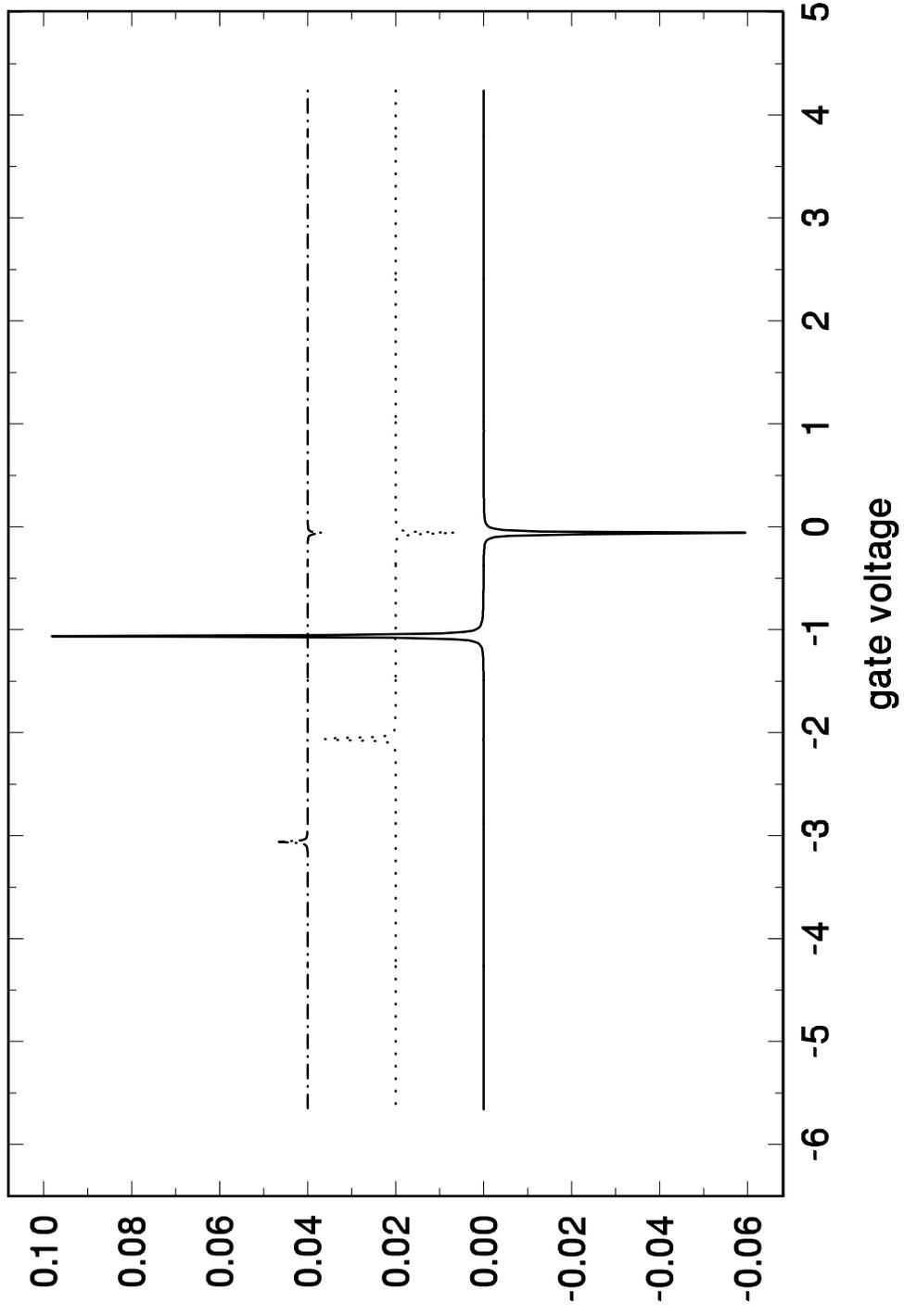
\begin{figure}
\caption{
The integrand of Eq.(23) versus gate voltage for $n=1$ (solid line), 
$n=2$ (dotted line), and $n=3$ (dash-dotted line). The system 
parameters are the same as that of Fig.1. For illustrating purpose, we have 
offset the curve by $0.02$. 
}
\end{figure}

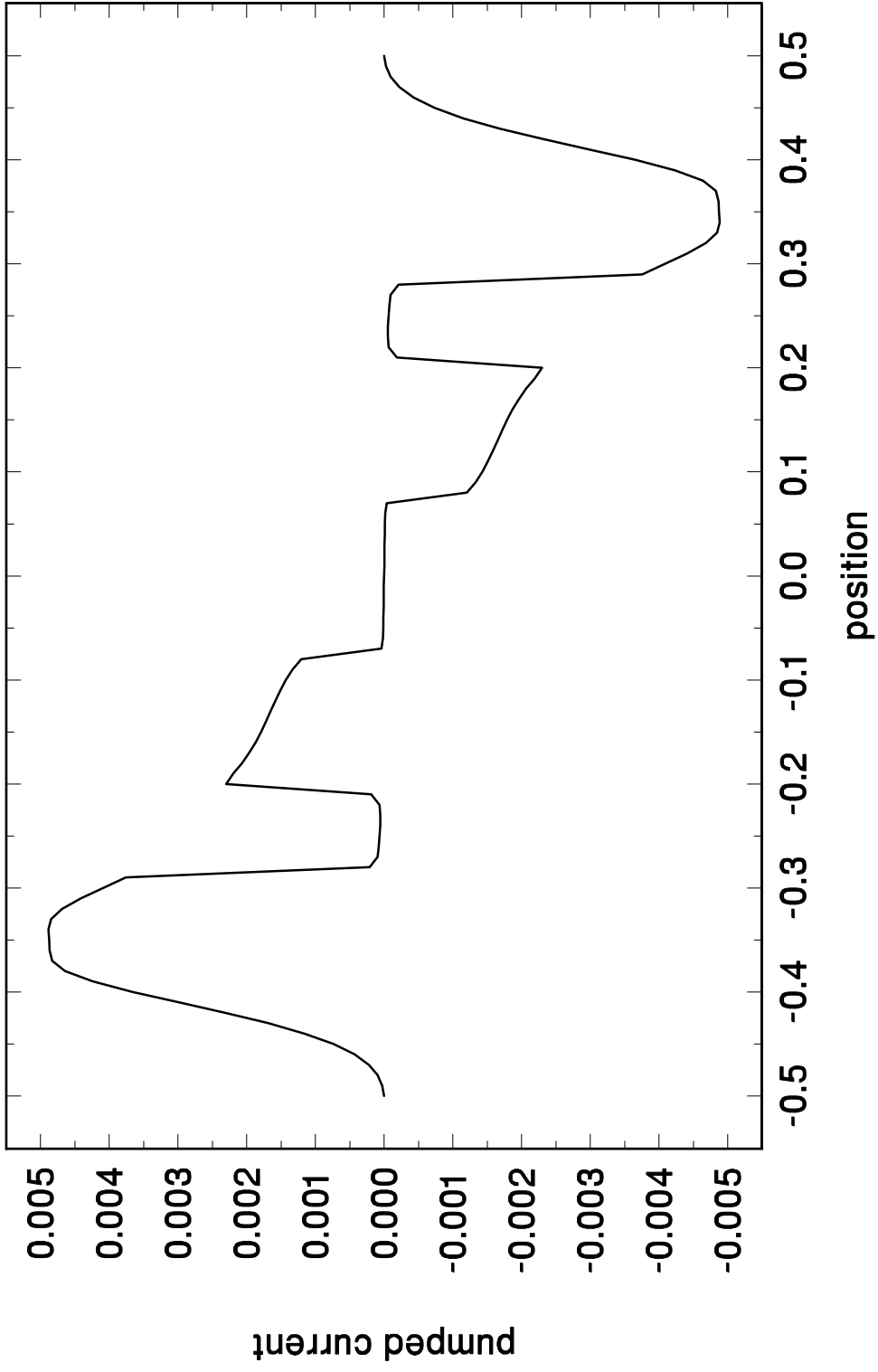
\begin{figure}
\caption{
The pumped current versus position of the pumping potential. Here
$v_g=-0.67$, $\omega=1$, and $v_p=6$. 
}
\end{figure}


\begin{thebibliography}{00}
\bibitem{foot1}
The result in Ref.\onlinecite{wbg1} has also been obtained using scattering 
matrix approach in Ref.\onlinecite{buttiker1}.
\bibitem{foot3}
Similar to Ref.\onlinecite{buttiker2}, we have used the adiabatic 
approximation so that the ${\cal G}^r(E+n\omega/2,T)= {\cal G}^r(E,T)$.
\bibitem{foot6}
Since $G^r_n$ is the renormalized Green's function, it also has the photon
emission component through the dependence of $\beta^r_{-n}$.
\end{thebibliography}
\end{document}